\documentstyle[twocolumn,aps,prl]{revtex}
\begin{document}

\draft

\twocolumn[\hsize\textwidth\columnwidth\hsize\csname @twocolumnfalse\endcsname

\title{Insights into molecular conduction from I-V asymmetry}

\bigskip

\author{A. W. Ghosh, F. Zahid, P. S. Damle and S. Datta}
\address{ School of Electrical and Computer Engineering, Purdue
University, W. Lafayette, IN 47907}%

\maketitle

\medskip

%\author{P. S. Damle, A. W. Ghosh and S. Datta}
%\address{ School of Electrical and Computer Engineering, Purdue
%University, W. Lafayette, IN 47907}%

\widetext
\begin{abstract}
We investigate the origin of asymmetry in the measured current-voltage (I-V) 
characteristics of molecules with no inherent spatial asymmetry. We
establish that such molecules can exhibit asymmetric I-V characteristics
due to unequal coupling with the contacts. In contrast with spatially asymmetric 
molecules, conduction takes place through essentially the same level in both
bias directions. The asymmetry arises from a subtle difference in the charging
effects, which can only be captured in a self-consistent model for molecular
conduction. For HOMO-based conduction, the current is smaller for positive voltage on
the stronger contact, while for LUMO conduction, the sense of asymmetry is 
switched.
\end{abstract}
\bigskip

\pacs{PACS numbers: 85.65.+h, 73.23.-b,31.15.Ar}
%31.15.Ar Ab initio calculations
%81.07.Nb Molecular Nanostructures
%81.07.Lk Nanocontacts
%85.65.+h Molecular electronic devices
%72.10.Bg General formulation of transport theory
%72.20.Dp General theory, scattering mechanisms of conductivity
%73.23.-b Electronic transport in mesoscopic systems
%73.40.Sx Metal-semiconductor-metal structures
%73.63.-b Electronic transport in mesoscopic or nanoscale materials and structures
%2col
%end of wide text
]
\narrowtext
%2col

Future electronic devices are quite likely to incorporate molecular
components, motivated by their size, mechanical flexibility and
chemical tunability. This vision is nearing reality with the capacity
to self-assemble, functionalize, and reproducibly measure the
current-voltage (I-V) characteristics of small groups of molecules.
Molecular I-Vs have revealed a wide range of conductance properties,
from metallic conduction in carbon nanotubes \cite{rNT} and quantum
point contacts \cite{rQPC}, to semiconducting behavior in DNA
\cite{rDNA} and conjugated aromatic thiols \cite{rReed1}, and
insulating behavior in alkyl thiol chains \cite{ralkane}.  Interesting
device characteristics such as rectification \cite{rMetzger}, switching
\cite{rReed2} and transistor action \cite{rSchon} have also been
reported.

The classic paradigm for asymmetry in molecular I-V measurements is the
Aviram-Ratner diode, consisting of a semi-insulating molecular species
bridging an electron donor-acceptor pair \cite{rAviram}. A positive
bias on the contact at the donor end brings the energy levels on the
donor and acceptor sites into resonance, while the opposite bias moves
the system away from resonance, leading to a strongly asymmetric I-V
characteristic \cite{rMetzger,rReed3}.  Spatial asymmetry in the
molecule is essential in generating the I-V asymmetry, causing the
energy levels, the electrostatic potential and electron wavefunctions
to be quite different for positive and negative voltages
\cite{rVuillame}.

In this paper, we address an I-V asymmetry experimentally observed for
spatially symmetric molecules \cite{rReichert,rKerg,rReifenberger} that
is qualitatively different from and weaker than the rectification in
spatially asymmetric molecules. The molecular I-V curves seen for these
systems start off being symmetric, but pick up a weak, reversible
asymmetry as the contacts are manipulated (Fig. 1). In contrast to
spatially asymmetric molecules, conduction in these molecules at
opposite voltages occur through essentially the same molecular levels
with very similar wave functions. We show that the origin of the
observed asymmetry is nontrivial, involving self-consistent shifts in
the energy levels due to charging effects. Asymmetry in charging arises
due to unequal coupling with the contacts, and can be present in
conduction measurements performed with a break junction
\cite{rReichert}, an STM tip \cite{rReifenberger} or an evaporated gold
contact \cite{rReed3}.  Remarkably, the sense of the asymmetry depends
crucially on whether the conducting level is HOMO or LUMO.  We
establish that {\it{ for a spatially symmetric molecule the current is
lower for positive bias on the stronger contact if conduction is
through a HOMO level, and higher if conduction is through a LUMO
level.}}

\begin{figure}
\vspace{1.7in}
\hskip 1cm\includegraphics{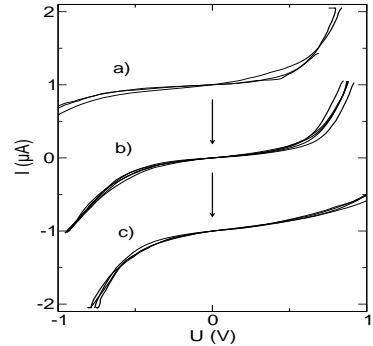}
\vskip 0.1cm
\caption{Sets of I-V characteristics (vertically shifted) observed by
Reichert et al.  \protect\cite{rReichert}  for a break junction
contacting a spatially symmetric molecule. To start with the I-V
characteristics are symmetric (center). On weakening one of the contacts
by pulling, the I-V curves become asymmetric. Two opposite sets of
asymmetric I-Vs are obtained with the same molecule simply by pulling on
the contacts (top and bottom).}
\label{f1} 
\end{figure}

Although ab-initio calculations exist for molecular conduction
\cite{rLang,rGuo,rDamle}, there is uncertainty about the
position of the Fermi level even in idealized metal-molecular
heterostructures. This uncertainty translates into a corresponding
ambiguity in the conductance gap in molecular I-Vs, as well as the
nature of the conducting orbitals. For phenyl dithiol (PDT) and its
derivative molecules coupled to gold (111) contacts, the Fermi level
has variously been suggested to be mid-gap \cite{rWenzel}, closer to
the HOMO level \cite{rDamle,rEmberly,rRatner}, or closer to  the LUMO
\cite{rLang,rHush}.  Although all these models predict similar
symmetric I-Vs, the sense of the I-V asymmetry for asymmetric contacts
pins down the nature of the conducting orbital rather
restrictively; for instance, the I-V asymmetry measured for PDT
with an STM tip \cite{rReifenberger} indicates HOMO-based 
conduction, as we show later. 

\begin{figure}
\vspace{2.1in}
\hskip 1.8cm\includegraphics{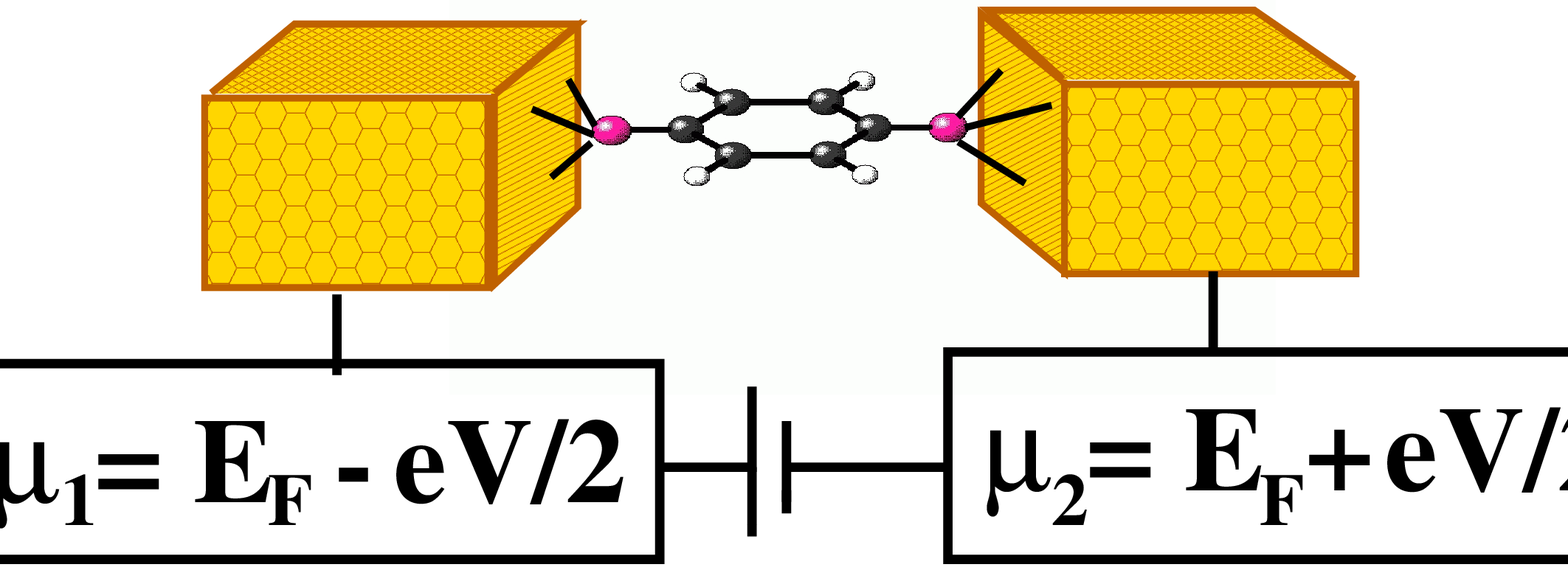}
\vskip 5.6cm
\hskip -0.3cm\includegraphics{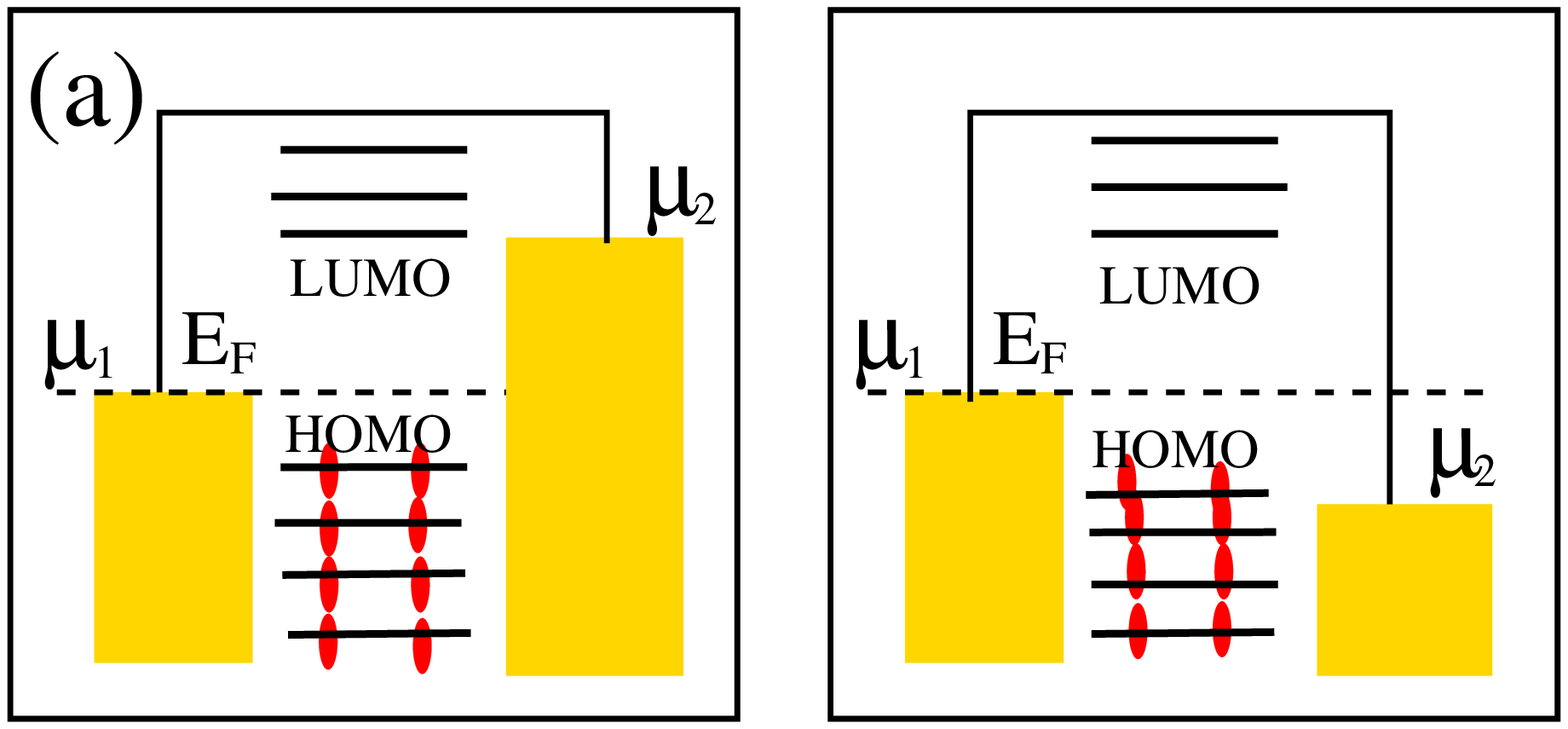}
\vskip 0.0cm
\hskip 0.5cm\includegraphics{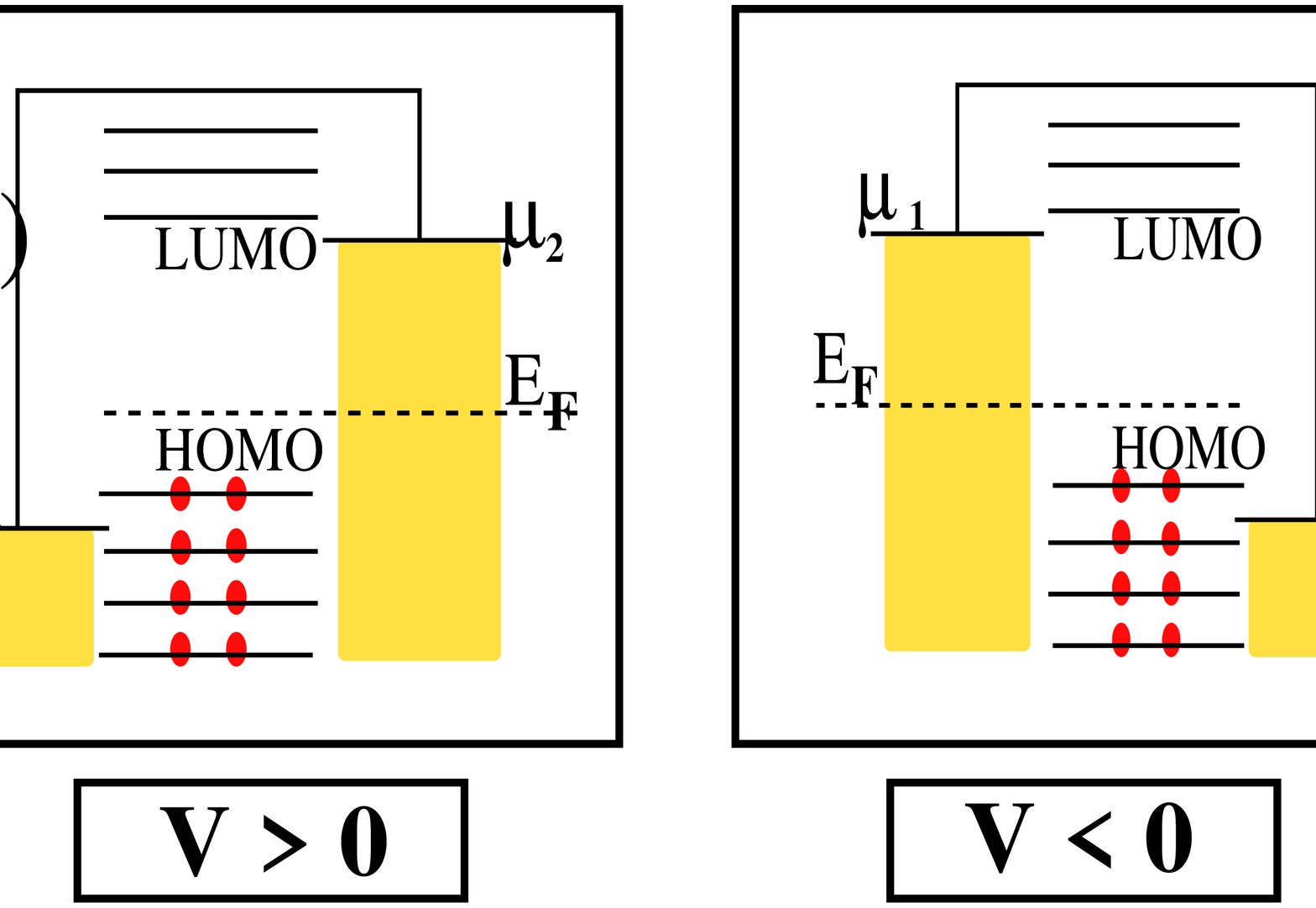}
\vskip -0.5cm
\caption{Schematic description of resonant current conduction
through discrete molecular levels broadened by interaction with the
contacts. Onset of current corresponds to crossing a molecular level by
a contact chemical potential under bias. For a molecule
strongly coupled to a substrate, and with a large gap DOS (a), the chemical
potential of the other (weaker) contact crosses the HOMO and LUMO
levels at opposite bias. In a typical molecular conductor (b) on the other hand,
the same level is crossed both ways.} 
\label{f2} 
\end{figure} 

{\it{Non self-consistent description of molecular conduction}}.  An
isolated molecule has discrete molecular levels, that are broadened
upon coupling with contacts into a continuous density of states (DOS)
due to hybridization with the metal wave functions. For molecules like
PDT the broadening is weak enough that the DOS shows a clear HOMO-LUMO
gap (HLG). The difference in the work functions between the metal and
the molecule leads to charge transfer between them, and the molecular
levels adjust due to Coulomb charging until the heterostructure is in
equilibrium. Under bias the contact chemical potentials split by the
applied voltage ($\mu_2 - \mu_1 = eV$). As long as both potentials lie
in the HLG, all HOMO levels are occupied and LUMO levels unoccupied,
and there is essentially zero current.  Once the bias is large enough
that a contact chemical potential crosses a molecular level, that level
is filled by one contact and emptied by the other, and starts
conducting current.  Thus the specific levels contributing to
conduction for either bias direction depend on the way $\mu_1$ and
$\mu_2$ are disposed relative to the molecular levels under bias.

One can get an asymmetric I-V with a symmetric molecule provided the
latter is coupled strongly enough to the substrate to be in equilibrium
with it, resulting in $\mu_1 = E_F$. The potential of the other contact
(say, an STM tip) moves freely under bias, $\mu_2 = E_F + eV$ (Fig.
2a). If furthermore there is a large density of metal-induced gap
states (MIGS) in the gap, then the electrostatic potential and the
molecular levels stay pinned to $\mu_1$ and don't vary with $V$. For
positive and negative substrate bias, $\mu_2$ crosses the LUMO and HOMO
levels respectively, yielding a strongly asymmetric I-V, since the LUMO
and HOMO transmissions are quite different in general.  The conductance
gap is then given by the HLG, a property intrinsic to the molecule
itself.

In practice, however, spatially symmetric molecular
species typically exhibit symmetric I-Vs\cite{rReed1}, with a
conductance gap that varies from experiment to experiment
\cite{rReed1,rSchon2}. The MIGS are usually not appreciable
causing the electrostatic and chemical potentials to separate.
The average electrostatic potential across the
molecule is $V/2$, which shifts the levels along with the applied bias.
Relative to the molecular levels therefore, the contact potentials move
in opposite directions symmetrically, $\mu_{1,2} = E_f \mp eV/2$ (Fig.
2b). {\it{The same molecular level (HOMO in the figure) is crossed for
either bias by the contact chemical potentials}}, leading to a
symmetric I-V, with a conductance gap given by $4(E_f - E_{\rm{HOMO}})$
\cite{rDatta}. The gap is no longer the HLG, but depends on both the
molecular chemistry and the contact microstructure.

The above non self-consistent description of molecular conduction can
be put on a quantitative footing by using an appropriate molecular Fock
matrix $F$ (ab-initio or semi-empirical), followed by a non-equilibrium
Green's function (NEGF) formulation of transport \cite{rDamle,rDatta2}.
For a given Fock matrix $F$, overlap matrix $S$, and contact
self-energies $\Sigma_{1,2}$ with corresponding broadenings
$\Gamma_{1,2} = i\left(\Sigma_{1,2}-\Sigma_{1,2}^\dagger\right)$, the
energy levels are given by the poles of the nonequilibrium Green's
function $G$, while their occupancies are obtained from the
corresponding density matrix $\rho$ and the contact Fermi functions
$f_{1,2}$:
\begin{eqnarray}
G(E) &=& \left(ES-F-\Sigma_1-\Sigma_2\right)^{-1}\nonumber\\
\rho &=& \left(1/2\pi\right)\int_{-\infty}^{\infty}dE\left(f_1G\Gamma_1 G^\dagger + 
f_2G\Gamma_2 G^\dagger\right)\nonumber\\
f_{1,2}(E) &=& \left[ 1 + \exp{((E - \mu_{1,2})/k_BT)} \right]^{-1}.
\label{e1}
\end{eqnarray}
The number of electrons $N$ and the steady-state current $I$ are then given by:
\begin{eqnarray}
N &=& 2~(\rm{for}~\rm{spin}) \times {\rm{trace}}(\rho~S) \nonumber\\
I &=& {{2e}\over{h}}\int_{-\infty}^\infty dE~{\rm{trace}} (\Gamma_1 G \Gamma_2 G^\dagger)\left[f_1(E)
- f_2(E)\right].
\label{e3}
\end{eqnarray}
\noindent The important point to notice is that since the same molecular level 
is being crossed both ways, the current is symmetric in the coupling
constants $\Gamma_{1,2}$ and one is stuck with a symmetric I-V for a symmetric
molecule, {\it{no matter how asymmetric the contacts are.}}

\begin{figure}
\vspace{2.0in}
\hskip -0.2cm\includegraphics{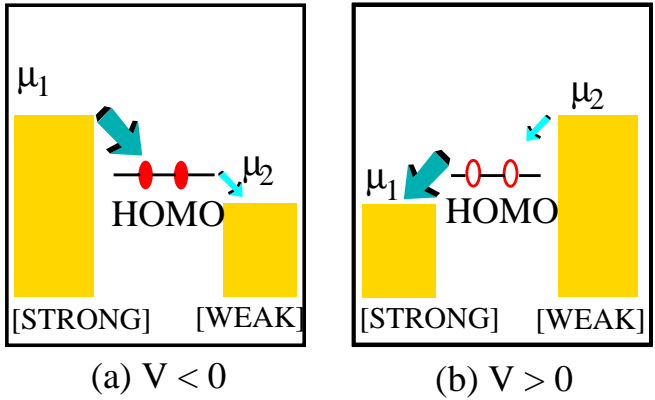}
\vskip 0.2cm
\caption{Origin of asymmetry in charging: one side (say a substrate) is
strongly contacted while the other is weak. Although the same level is
crossed by the contact potentials for opposite bias, for positive substrate
voltage (b) the HOMO level is emptied by the stronger contact, which
positively charges the molecule and shifts the energy levels down. Such a
shift, not present for negative substrate bias (a), postpones the onset of
conduction and effectively stretches out the voltage axis in the direction of
positive substrate bias. }
\label{f3}
\end{figure}

The asymmetric I-V cannot simply be explained by suggesting that the
average electrostatic potential across the molecule is $\eta V$
\cite{rDatta}, $\eta$ decreasing from 0.5 as a contact is drawn
out.  Such a theory yields an asymmetry only in the onset voltages for
conduction, rather than a dragged out I-V as in Fig. 1 (the latter actually
corresponds to different conductance peak values in the two bias
directions).  One will then need to invoke a voltage-dependent $\eta$ to 
explain Fig.1, which formally amounts to our self-consistent charging 
description below.

{\it{Self-consistent description (Charging induced asymmetry)}}
The above description of molecular conduction does not include the effects of
charging and electronic Stark shift back on the levels. For a small molecule
with a large ($\sim 1$ eV) capacitative charging energy, charging effects need
to be taken into account self-consistently. In the following, we show
that such effects can lead to contact-induced asymmetries of the
sort observed in Fig. 1.

The effect of differential charging on the I-V asymmetry is explained
schematically in Fig. 3 for a single level. Let us suppose that the
contacts are strongly asymmetric ($\Gamma_1 \gg \Gamma_2$). For
negative bias on the strong contact (a), the latter is trying to fill
the nearest (HOMO) level, while the weaker contact is trying to empty
it, with the net result that the HOMO level stays filled, with current
onset set by the voltage where $\mu_2$ first crosses the (neutral)
molecular level.  For positive bias (b), however, the HOMO level is
emptied out, which charges up the molecule positively. This adds a
self-consistent charging energy that lowers all the energy levels,
postponing thereby the point where the HOMO is crossed by $\mu_1$. In
effect, this stretches out the voltage axis, leading thus to a smaller
conductance for positive bias on the stronger contact, as mentioned earlier.  For
LUMO-based conduction the argument is reversed, since filling the LUMO
level charges up the molecule negatively.

\begin{figure}
\vspace{3.3in}
\hskip -1cm\includegraphics{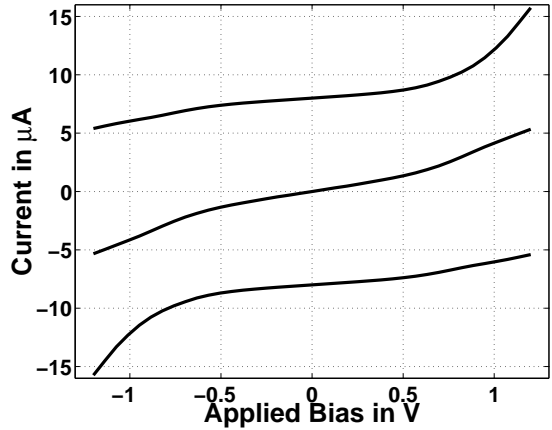}
\vskip -2.2cm
\caption{Extended H\"uckel-based I-V characteristics for Phenyl Dithiol
(PDT), including self-consistent charging effects. The central graph
(symmetric) is for symmetric weak couplings ($\Gamma$s decreased symmetrically
 by a factor of 10 relative to their chemisorbed values) of the PDT molecule
with Au(111) contacts. The lower graph corresponds to decreasing
$\Gamma_2$ five times relative to $\Gamma_1$, which stretches out the
voltage axis in the direction of the stronger contact, i.e., along
positive bias on contact 1. Reversing the couplings yields the upper
graph. This graph can also be obtained without reversing the couplings,
by simply raising $E_f$ so that the conduction occurs through the LUMO
level instead.} 
\label{f4} 
\end{figure}

We illustrate self-consistent charging effects in a PDT molecule
modeled by a extended H\"uckel \cite{rDatta} description of the Fock
matrix $F$, coupled to the NEGF equations  described earlier. The
self-energies are calculated for Au(111) contacts using a recursive
technique \cite{rDamle}. Charging effects are incorporated in $F$ by
adding a term $U_{\rm{SC}}$ describing the deviation from the
equilibrium electron count $N_0$ within the local density approximation
\cite{rabinitio}
\begin{equation}
U_{\rm{SC}} = U_0\left(N - N_0\right),
\label{e4}
\end{equation}
and solving for $F$ self-consistently with equations \ref{e1} and
\ref{e3}.  The constant charging energy $U_0$ ($\sim 0.7$eV per
electron) corresponds to a flat electrostatic potential profile in the
molecule \cite{rDatta}, and serves to shift all the energy levels
rigidly by a constant.

Fig. 4 shows the I-V characteristic for PDT with weak symmetric
contacts (central graph), with contact self-energies 0.1 times their
chemisorbed values, assuming HOMO conduction ($E_f = 10$ eV). Weakening
the coupling with the right contact ($\Gamma_2 = 0.2~
\Gamma_1$) stretches the voltage axis along contact 1 (lower
figure).  The I-V asymmetry can be switched (upper figure) either by
switching the contact couplings, or by
raising $E_f$ enough to generate LUMO-based conduction.

The calculated trend in the I-V characteristics agrees qualitatively
with the experiments on symmetric molecules with asymmetric contacts
\cite{rReichert,rKerg,rReifenberger}.  In break-junction experiments in
general, the molecules are likely to adsorb separately onto the two
contacts symmetrically \cite{rKircz}, leading to symmetric I-Vs
\cite{rReed1} even for asymmetric contacts. In carefully controlled
break-junction experiments, however, a single molecule is expected to
couple simultaneously with both contacts. This allows one to induce
asymmetry easily by manipulating the contacts, changing the I-V from
symmetric to asymmetric and back \cite{rReichert}, as well as flipping
the sense of the asymmetry by interchanging contact couplings. Weak I-V
asymmetries are also seen in STM measurements of symmetric molecules
such as PDT \cite{rReifenberger}, where the conductance is lower for
positive substrate bias. Since the coupling of the molecule with the
STM contact is evidently weaker than with the substrate, the above
asymmetry implies that conduction in PDT is HOMO-based.  
%Explanations
%of the I-V %asymmetry in terms of tunneling barrier asymmetries or MIGS
%\cite{rMIGS} %usually predicts much stronger asymmetries than are
%typically observed.

In summary, we have established that even for spatially symmetric
molecules, contact asymmetry can induce an asymmetric I-V through
differential charging, the sense of the asymmetry depending on whether
conduction is through a HOMO or a LUMO level.

We would like to thank M. Paulsson, R. Reifenberger, D. Janes and H. B. Weber
for useful discussions. This work has been supported by NSF and the
US Army Research Office (ARO) under grants number 9809520-ECS and
DAAD19-99-1-0198.


\begin{thebibliography}{100}
\bibitem{rabinitio} A rigorous theory \cite{rDamle} replaces the extended
H\"uckel Fock matrix, including the self-consistent charging term, by a proper 
DFT-based Fock matrix. However, a spatially constant potential profile \cite{rDatta}
with an adjustable $E_f$ matches symmetric I-V characteristics quantitatively, and
can be attributed to screening of the electrostatic potential in the molecule.
\end{thebibliography}
\end{document}